\documentclass[aps,showpacs,showkeys]{revtex4}
\usepackage{amsmath, amsthm, amssymb,epsfig,alltt}

\newcommand{\nn}{\nonumber }

\def\a{\alpha}

\def\e{\epsilon}
\def\p{\partial}
\def\m{\mu}
\def\n{\nu}
\def\t{\tau}
\def\th{\theta}
\def\s{\sigma}
\def\g{\gamma}

\def\half{\frac{1}{2}}

\def\barz{{\bar z}}

\def\sp{\sigma^\prime}
\def\nn{\nonumber}

\def\2pap{2\pi\alpha^\prime}

\def\beq{\begin{eqnarray}}
 \def\eeq{\end{eqnarray}}
 \def\4pap{4\pi\a^\prime}
 
 \def\sp{{\s^\prime}}
 
 \def\ap{{\a^\prime}}

 \def\barz{{\bar z}}

\begin{document}


\title[Short Title]{Boundary State for Dissipative Quantum Mechanics 
 and Thirring Model}

\author{Taejin Lee}
\email{taejin@kangwon.ac.kr}

\affiliation{
Department of Physics, Kangwon National University,
Chuncheon 200-701, Korea}

\affiliation{
Pacific
Institute for Theoretical Physics Department of Physics and
Astronomy, University of British Columbia, 6224 Agricultural
Road, Vancouver, British Columbia V6T 1Z1, Canada}

\date{\today}

\begin{abstract}
The dissipative quantum system is studied using the Thirring model with a boundary mass. At the critical point where the Thirring coupling vanishes, the theory reduces to a free fermion theory with a boundary mass. 
We construct boundary states for the dissipative quantum systems in one dimension, which describes the system off the critical points as well as at the critical points. 
\end{abstract}

\maketitle

\section{Introduction}

The two space-time dimensional quantum field theory of massless
bosons with a periodic boundary potential has recently come into the
spotlight again. This boundary conformal field theory has received 
constant attention along the years, since it describes 
the various condensed physics systems such as
the dissipative quantum mechanics of a particle in a one-dimensional periodic potential \cite{schmid,Guinea,fisher,Callan:1989mm}, 
Josephson junction arrays \cite{larkin,fazio,sodano} and the
dissipative Hofstadter problem \cite{Callan:1991da,Callan:1992vy}.
The application of the theory also includes the Kondo problem
\cite{Affleck:1990iv,Affleck:1990by}, the one-dimensional
conductors \cite{Kane:1992}, tunnelling between Hall edge states
\cite{kane2}, and junctions of quantum wires~\cite{Oshikawa:2005fh}.
The recent revival of interest in the theory is mainly due to the
string theory application. The string with its ends on a unstable D-brane develops a marginal periodic boundary interaction and the tachyon field of the open string condensates \cite{Witten:1992qy,Shatashvili:1993kk,Shatashvili:1993ps,Tlee:01nc,Tlee:01os,Tlee:01one}. This process called ``rolling tachyon" \cite{Sen:2002nu,Tlee:06} is believed to be responsible for the decay of the unstable D-brane. 

Recently the theory with a periodic boundary potential is discussed in detail
by applying the fermionization technique \cite{Tlee:2005ge,Hassel}. The advantage of the fermion formulation is that one can explicitly construct the boundary state for the rolling tachyon, since the periodic boundary potential becomes a boundary fermion mass term, which is quadratic in fermion field.
The fermion formulation of the theory, however, has been performed only for the theory at the critical point. In this paper we shall generalize the fermion formulation to the theory off the critical point and develop a perturbative theory around the critical point by using the Thirring model. 

The massless Thirring model \cite{Thirring}, which is the first example of an exactly solvable relativistic interacting field theory, has served as an excellent laboratory for the study of various aspects of the quantum field theory in two dimensions. Since the seminal work of Thirring, extensive studies of the model have been carried out by numerous authors \cite{Glaser,Johnson,Hagen,Klaiber}; notably
the complete solution of the theory was obtained by Klaiber \cite{Klaiber}.
Along this line Schwinger \cite{Schwinger} obtained an exact solution of quantum electrodynamics in 1+1 dimensions and Coleman \cite{Coleman} proved the equivalence between the sine-Gordon model and the massive Thirring model.

In this Letter we will widen the range of applications of the Thirring model
by describing the dissipative quantum system in terms of the Thirring model.
We find that the Thirring model with a boundary mass is the most suitable 
framework to discuss the dissipative quantum system: At the 
critical point the Thirring coupling,
which is directly related to the friction constant of the system,
vanishes and the theory reduces to a free fermion theory with a boundary mass. The boundary state takes a simple form at the 
critical point when it is expressed in fermion variables
as discussed in refs. \cite{Tlee:2005ge,Hassel}. Thus, the Thirring model provides a perturbative theory expanded in the Thirring coupling near the critical point. 
The previous analysis of the dissipative quantum system is based on this Coulomb gas expansion; the duality symmetries and the phase diagram of the system have been studied in this framework.
However, the perturbative theory based on the Coulomb gas expansion  
has some limitation to study the system near or off the critical point, being one dimensional theory with a non-local interaction. 
In the followings we formulate the dissipative system of a particle
moving in one dimension subject to a periodic potential, known as 
the Schmid model, near the critical point in terms of the Thirring model and the boundary state formulation.



\section{The Schmid Model and The Thirring Model}

Caldeira and Leggett \cite{caldeira83ann,caldeira83phy} discussed first
the quantum mechanical description of dissipation by coupling a bath or environment, which consists of an infinite number of harmonic oscillators, to the system. 
In the quantum theory the interaction with the bath produces a non-local effective interaction. Subsequenlty Schmid \cite{schmid} studied the dissipative system in the presence of a periodic potential. The one dimensional dissipative model with a periodic potential, called Schmid model, is described by the following action
\beq 
S_{SM} &=& \frac{\eta}{4\pi \hbar} \int^{T/2}_{-T/2} dt dt^\prime 
\frac{\left(X(t) - X(t^\prime)\right)^2}{(t-t^\prime)^2} 
- \frac{V_0}{\hbar} \int^{T/2}_{-T/2} dt \cos \frac{2\pi X}{a}. 
\eeq
The first non-local term is responsible for the dissipation, 
and the second term denotes the periodic potential respectively. An interesting feature of the model is that it exhibits a phase transition, unlike one dimensional quantum mechanical systems with local interactions only. Depending on the value of the friction constant $\eta$, the system has two phases; the localized phase and the delocalized one. 

We can map the Schmid model to the string theory on a disk
by identifying the time as the boundary parameter $\sigma$ 
in string theory and scaling the field variable $X$:
\beq
t = \frac{T}{2\pi} \tau, \quad X \rightarrow \frac{a}{2\pi} X.
\eeq
Then, the action for the Schmid model reads as 
\beq
S_{SM} &=& \frac{\eta}{4\pi \hbar} \left(\frac{a}{2\pi}\right)^2 
\int d\tau d\tau^\prime \,
\frac{\left(X(\t) - X(\t^\prime)\right)^2}{(\t-\t^\prime)^2} 
- \frac{V_0}{\hbar} \frac{T}{2\pi} \int d\tau \,\half
\left(e^{iX} + e^{-iX} \right). 
\eeq
This action can be interpreted as the boundary effective action 
for the open bosonic string subject to a boundary periodic potential on a disk with a boundary condition; on the boundary
$\p M$, $X(\s,\t) = X(\t)$,
\beq
e^{-iS_{SM}} &=& \int D[X] \exp\Biggl[ - i\left(\frac{1}{4\pi \ap}
\int_M d\t d\s \p_\a X \p^\a X  - \frac{m}{2}
\int_{\p M} d\t \left(e^{iX}+ e^{-iX} \right)\right)\Biggr]. 
\eeq 
Here, we identify the physical parameters of the two theories as 
\beq
\frac{\eta}{4\pi \hbar} \left(\frac{a}{2\pi}\right)^2 
= \frac{1}{8\pi^2 \ap}, \quad 
-\frac{V_0}{\hbar} \frac{T}{2\pi} = m.
\eeq
In string theory the periodic potential describes the interaction
between the open string the unstable D-brane. 

The open string dynamics is often described more efficiently in its equivalent closed string picture by the boundary state formulation. The corresponding closed string action is obtained from its open string action by simply taking $\s \rightarrow \t$, $\t \rightarrow \s$,
\beq
S = \frac{1}{4\pi \ap}
\int d\t d\s\, \p_\a X \p^\a X  - \frac{m}{2}
\int d\s \left(e^{iX}+ e^{-iX} \right)
\eeq
The open string dynamics can be encoded completely by the boundary state.

When $\a= 1/\ap= 1$, the system becomes critical. This can be
easily understood if we introduce an auxiliary boson field $Y$
and fermionize the system \cite{Tlee:2005ge,Hassel,Polchinski:1994my}. 
Introducing an auxiliary boson field
$Y$ which satisfies the Dirichlet condition $Y|_{\p M} = 0$ at the
boundary, and defining the boson fields,
$\phi_1 = \frac{X+Y}{\sqrt{2}}, ~~~
\phi_2 = \frac{X-Y}{\sqrt{2}}$, 
we may rewrite the action as
\beq
S =  \frac{\a}{4\pi}\int_M d\t d\s \sum_i^2 \p \phi_i \p \phi_i
- \frac{m}{4} \int_{\p M} d\t \sum_i^2 \left(e^{i\sqrt{2}\phi_i}+ e^{-i\sqrt{2} \phi_i}\right).
\eeq
Since $Y$ is a free boson field and it vanishes at the boundary,
$Y$ is completely decoupled from the physical degrees of freedom. 
We see that if $\a=1$, 
$e^{\pm i\sqrt{2} \phi_i}$ are marginal boundary operator with the scaling dimension $1$. 
An explicit calculation of the current correlation function or the mobility shows that 
\beq
\langle 0| \p_\s X(\s) \p_\s X(\sp) |B \rangle = - \frac{1}{2} (1-\pi^2 m^2) 
\sin^{-2}\frac{(\s-\sp)}{2}
\eeq
the theory becomes indeed critical where $\a =1$.

An explicit evaluation of the boundary state at the critical point is possible if we fermionize the model; the boson fields are mapped to the fermion fields as 
\begin{subequations}
\label{generallabel}
\begin{eqnarray}\label{fermionization1}
\psi_{1L}(z)&=&\zeta_{1L}:e^{-\sqrt{2}i\phi_{1L}(z)}:,~~
\psi_{2L}(z)=\zeta_{2L}:e^{\sqrt{2}i\phi_{2L}(z)}: \\
\psi_{1R}(\bar z)&=&\zeta_{1R}:e^{\sqrt{2}i\phi_{1R}(\bar z)}:,~~
\psi_{2R}(\bar z) = \zeta_{2R} :e^{-\sqrt{2}i\phi_{2R}(\bar z)}:
\label{fermionization2}
\end{eqnarray}
\end{subequations}
where $\zeta_{iL/R}$ are co-cycles, ensuring the anti-commutation
relations between the fermion operators. Since the boundary interaction term can be written as a boundary fermion mass term, which is only quadratic in fermion field, the model is exactly solvable
\beq
S &=& \int \frac{d\tau d\sigma}{2\pi} ~
\left(\bar\psi_1 \g^\m \p_\m \psi_1 + \bar\psi_2 \g^\m \p_\m \psi_2\right) + m \int \frac{d\s}{2\pi} \left(\bar\psi_1 \psi_1
+ \bar\psi_2 \psi_2 \right)
\eeq
where $\psi_i = (\psi_{iL}, \psi_{iR})^t$, and
\beq
\gamma^0 &=& \s_1, \quad
\gamma^1 = \s_2, \quad 
\gamma^5 = \s_3 = -i \gamma^0 \gamma^1. 
\eeq
The boundary state is given formally as
\beq
|B \rangle = :\exp\left[m \int \frac{d\s}{2\pi} \left(\bar\psi_1 \psi_1 + \bar\psi_2 \psi_2 \right)\right]:|N,D\rangle
\eeq
where $|N,D\rangle$ is a simple boundary state satisfying
\beq \label{simple}
\left(\psi_R(0,\s) +i \s^2 \psi_L(0,\s)\right)|N,D\rangle = 0,~~
\left(\psi^\dagger_R(0,\s) +i \psi^\dagger_L(0,\s)\s^2\right)|N,D\rangle =0.
\eeq
We refer the reader to ref. \cite{Hassel} for the explicit expression of the boundary state. 

Now let us discuss the dissipative system off the critical points. When $\a \not =1$, we 
may write the action as
\beq
S &=&  \frac{1}{4\pi}\int_M d\t d\s \sum_i^2 \p \phi_i \p \phi_i
+ \frac{1}{4\pi}\left(\a -1 \right)\int_M d\t d\s \sum_i^2 \p \phi_i \p \phi_i \nn\\
&& - \frac{m}{4} \int_{\p M} d\t \sum_i^2 \left(e^{i\sqrt{2}\phi_i}+ e^{-i\sqrt{2} \phi_i}\right).
\eeq
and treat the second term as an interaction. In terms of the fermion fields the second term can be written as the Thirring interaction term.
Hence, the fermionized action is given by
\beq
S &=&  \frac{1}{2\pi} \int_M d\t d\s \sum_i^2
\left(\bar{\psi}_i \gamma^\m \p_\m 
\psi_i + \frac{g}{4\pi} j^\m_i j_{i\m}\right) 
+ \frac{m}{2} \int_{\p M} d\s \sum_i^2 {\bar \psi}_i \psi_i
\eeq
where $g = \pi (\a -1)$.
This is the Thirring model with a boundary mass. At the critical point where $g =0$ ($\a = 1$), the action reduces to the free fermion theory with a boundary mass. Near the critical point, we can use this Thirring action to develop perturbative
theory for the dissipative quantum system. 


\section{Boundary State near the Critical Point}

In order to apply the boundary state formulation to the
dissipative system near the critical point, the bulk action 
should be free. We may transmute the bulk Thirring interaction
into a boundary one by introducing Abelian gauge fields
\beq
S &=& \frac{1}{2\pi} \int_M d\t d\s\sum_i^2
\left[
\bar{\psi}_i\gamma^\m\left(\p_\m + iA_{i\m}\right)\psi_i 
+ \frac{\pi}{g} A_{i\m} A_i{}^\m \right] 
+  \frac{m}{2} \int_{\p M} d\s \sum_i^2 {\bar \psi}_i \psi_i .
\eeq
Since in general the Abelian gauge vector fields in $1+1$ dimensions may be decomposed as 
\beq
A^\m_i = \e^{\m\n} \p_\n \th_i + \p^\m \chi_i, ~~~ i =1,2 ,
\eeq
the interaction between the gauge fields and the fermion fields
may be removed by a gauge transformation
\beq
\psi_i = e^{-\g_5 \th_i -i\chi_i} \psi_{i\,0},\quad 
\bar\psi_i = \bar\psi_{i\,0} e^{-\g_5 \th_i +i\chi_i}.
\eeq
Then the bulk action becomes a free field one
\beq
S_{bulk} = \frac{1}{2\pi} \int_M d\t d\s \sum_{i=1}^2
\left[
\bar\psi_{i\,0} \gamma^\m \p_\m \psi_{i\,0} + \left(\frac{\pi}{g}
+1\right)\left(\p\th_i\right)^2 + \frac{\pi}{g}
\left(\p\chi_i\right)^2\right].
\eeq
The additional kinetic action for $\th_i$ is a manifestation of 
the $U(1)$ chiral anomaly: 
\beq
D[\psi]D[\bar\psi] = D[\psi_0] D[\bar\psi_0] \exp
\left[\frac{1}{4\pi} \int d\t d\s \sum_i (\p \th_i)^2 \right].
\eeq
Since the boundary mass term is not invariant under the $U(1)$ chiral gauge transformation, it transforms as 
\beq
\sum_i \bar\psi_i\psi_i = \sum_i \bar\psi_{i0} e^{-2\g_5 \th_i} \psi_{i0}.
\eeq
Note that scalar fields $\chi_i$ are free in the bulk and do not appear in the boundary action. Since the physical operators, being 
$U(1)_V$ gauge invariant, do not depend on $\chi_i$, we may drop
them. For the sake of convenience, we scale the scalar fields
\beq
\th_i \rightarrow \kappa \th_i, ~~~
\kappa = \sqrt{\frac{g}{2(\pi+g)}}.
\eeq
It brings us to 
\beq
S = \frac{1}{2\pi} \int_M d\t d\s \sum_{i=1}^2
\left[
\bar\psi_{i\,0} \gamma^\m \p_\m \psi_{i\,0} + \half\left(\p\th_i\right)^2 \right] + m \int_{\p M} d\s \sum_i \bar\psi_{i0} e^{-2\g_5 \kappa\th_i} \psi_{i0}.
\eeq
It is clear that in the limit of the critical point, $\kappa \rightarrow 0$, the interaction between the scalar fields
$\th_i$ and $\psi_0$ vanishes. Thus, $\th_i$, becoming free 
fields, can be dropped and the action reduces to one at the
critical point. 



The next step to construct the perturbative boundary state formulation is to find appropriate boundary conditions for 
$\psi^i_0$ and $\th_i$. We note that the fermion fields $\psi^i_0$
should be understood as those at the critical point. So the boundary conditions for them are the same as those in Eq.(\ref{simple})
The boundary conditions for the fermion fields would remain intact, since the boundary conditions for the boson fields $\phi_i$ would not change if the Thirring interaction term is turned on. The boundary conditions for the fermion fields can be formally kept unchanged, if we require the boson fields satisfy the following conditions
\beq \label{boundary2}
\th_1 |B_0\rangle = -\th_2 |B_0 \rangle,~~
\chi_1 |B_0\rangle = \chi_2 |B_0 \rangle .
\eeq
As we already noticed that $\chi_i$ are completely decoupled 
from the physical degrees of freedom, we are not concerned with
the boundary conditions for $\chi_i$. Eq.(\ref{boundary2}) only fixes the boundary condition for $\frac{1}{\sqrt{2}}(\th_1 + \th_2)$.
For the sake of completness, we may choose a Neumann conditions for 
$\frac{1}{\sqrt{2}}(\th_1 - \th_2)$.
Once, the simple boundary state $|B_0\rangle$ is constructed,
the boundary state for the dissipated system $|B(m,\kappa)\rangle$ may be
given as 
\beq
|B(m,\kappa) \rangle = \exp\left[m \int_{\p M} d\s \sum_i \bar\psi_{i} e^{-2\g_5 \kappa\th_i} \psi_{i} \right]|B_0\rangle
\eeq
where we drop the subscript $``\,{}_0\,"$ of the fermion fields for notational convenience. 

It is interesting to see that the scalar fields $\th_i$ appear
only through the boundary interaction and the physical operators
such as currents do not depend upon them.
Near the critical point where $|\kappa| < 1$, we may expand the boundary state $|B(m,\kappa)\rangle$ in $\kappa$ as follows
\beq
|B(m,\kappa)\rangle &=& \exp\left[m\int d\s \sum_i \bar\psi_i \psi_i 
-2m\kappa \int d\s \sum_i \bar\psi_i \g^5 \psi_i \th_i \right]|B(0,0)\rangle \nn\\
&=& \sum_n \frac{(-2m\kappa)^n}{n!}\left[\int d\s \sum_i \bar\psi_i \g^5 \psi_i \th_i \right]^n |B(m,0)\rangle 
\eeq
where $|B(m,0)\rangle$ corresponds to the boundary state at the critical point, of which explicit expression can be found in \cite{Hassel}.

As we expand $|B(m,\kappa)\rangle$ in $\kappa$, we encounter
a divergent term, proportional to the boundary mass term
 with a coefficient
\beq
&& ~ 2m^2 \kappa^2 \int d\s_1 \int d\s_2 \sum_i \langle \th_i(\s_1) \th_i(\s_2)\rangle \Bigl( \langle \bar \psi \g^5 \psi (\s_1) 
\bar \psi \g^5 \psi (\s_2) \rangle \nn\\
&& = 2m^2 \kappa^2 \int d\s_1 \int d\s_2 \, \ln \left\vert
1 - \frac{z_1}{z_2} \right\vert \left( \frac{1}{z_1 - z_2} 
\frac{1}{\barz_1 - \barz_2} \right)
\eeq
where $z_i = e^{i\s_i}$ and $\barz_i = e^{-i\s_i}$.
In order to regularize it we may introduce an infinitesimal parameter
$|\e| \ll 1 $ as follows
\beq
z_1 = e^{i\s_1}, ~~~ \barz_1 = \frac{1}{z}_1, ~~~
z_2 = e^{-\e} e^{i\s_2}, ~~~ \barz_2 = \frac{e^{-2\e}}{z_2}.
\eeq
Then we find,
\beq
\int d\s_1 \int d\s_2 \, \ln \left\vert
1 - \frac{z_1}{z_2} \right\vert \left( \frac{1}{z_1 - z_2} 
\frac{1}{\barz_1 - \barz_2} \right)
=  \frac{\pi^2}{2\e}. 
\eeq
This divergence can be taken care of by renormalization of the boundary mass. 
It yields 
\beq
m = m_0 \left[1 + \frac{\a-1}{2\a} \ln \frac{\Lambda^2}{\mu^2}\right]=
m_0 \left(\frac{\Lambda^2}{\mu^2}\right)^{\frac{(\a-1)}{2\a}}
\eeq
where $\ln \frac{\Lambda^2}{\mu^2} = \frac{m^2_0 \pi^2}{\e}$.
The result is in complete agreement with the previous work ref.\cite{fisher}: 
If $\a >1$, $m$ tends to grow and if $\a <1$, it scales to zero. The corection
to $m$ vanishes when $\a =1$. In fact, an explicit construction of the boundary state showes that perturbative corrections to $m$ vanish at all orders if $\a=1$.

A few remarks are in order to conclude this Letter. We discuss the dissipative quantum system in one dimension, employing the boundary state formulation of string theory and the Thirring model in two dimensions. The framework presented here has some advantages over the previous one based on the Coulomb gas expansion which deals with the non-local interaction: We need to deal with a local boundary interaction only and can take an advantage of the string theory techniques to explore various aspects of the system. Since the system is described in terms of the (bulk) massless Thirring model, all the physical quantities are exactly calculable at any given order.

 \noindent {\bf Acknowledgement:} The author thanks Gordon Semenoff, Philip Stamp for informative discussions.

\end{document}